# Mechanical analogies for the Lorentz gauge, particles and antiparticles

Valery P. Dmitriyev

*Lomonosov University*
*P.O.Box 160, Moscow 117574, Russia*
*e-mail: dmitr@cc.nifhi.ac.ru*

**Abstract**

An exact analogy of electromagnetic fields and particles can be found in mechanics of a turbulent ideal fluid with voids. The system is supposed to form a fine dispersion of voids in the fluid. This microscopically discontinuous medium is treated as a continuum. The turbulence is described in terms of the Reynolds stresses. Perturbations of the homogeneous isotropic turbulence are considered. For the high-energy low-pressure turbulence they are usually small. This entails the linearization of the Reynolds equations. The latter appear to be isomorphic to Maxwell's electromagnetic equations. The Lorentz gauge expresses the slight effective compressibility of the medium. A particle can be viewed as a cavity in the medium. A respective antiparticle is modeled with an agglomerate of the medium's material. Microscopically, these correspond to some nonlinear vortex formations in the "vortex sponge".

## 1 Introduction

We are in search of a mechanical medium capable to reproduce or imitate the world of particles and physical fields. Earlier, a mechanical model of particles and fields has been proposed, which is based on the approximation of an incompressible substratum. Average turbulence in an ideal fluid was considered. In the ground state, the turbulence was taken to be homogeneous and isotropic. Perturbations of the background turbulence model the physical fields [1]. Voids in the turbulent fluid give rise to the structures, which can be taken as the model of the particles [2]. The condition of the substratum incompressibility manifests itself in the classical electromagnetism as the Coulomb gauge.

Now, we give some refinement of the above model. It is suggested that the substratum is represented by a volume distribution of the empty space in the ideal fluid. Microscopically, this is conveniently viewed as the vortex sponge – the plenum of hollow vortex tubes, which pierce the ideal fluid in all directions [3]. Then, the particles should be modeled by some nonlinear vortex structures e.g. by a loop on a vortex filament [4]. Further, such a rather complicated system will be described in a continuous approximation. We will take the substratum as a turbulent fluid continuum with the variable volume density $V(\mathbf{x},t)$. The particles will be modeled by discontinuities of the medium. The substratum compressibility manifests itself as the Lorentz gauge. The possibility for the medium density to have both positive and negative deviations from the background value enables us to model the symmetry in the particle-antiparticle pairs.

## 2 Turbulent substratum

We consider a fluid, which moves disorderly. In developed turbulence, the fluid velocity $\mathbf{u}(\mathbf{x},t)$ and pressure $p(\mathbf{x},t)$ can be treated as random variables. Following well known in hydrodynamics Reynolds scheme, we decompose them into the average and pulsation components, respectively:

$$\mathbf{u}=\langle\mathbf{u}\rangle+\mathbf{u}', \quad p=\langle p\rangle+p'. \qquad (2.1)$$

The turbulence can be characterized on the average by the set of consecutive moments $\langle u_i \rangle$, $\langle p \rangle$, $\langle u'_i u'_k \rangle$ and so on. The second moments $\langle u'_i u'_k \rangle$ have the meaning of the average turbulence energy density

$$1/2 V \langle u'_i u'_i \rangle.$$

In the ground state and also at infinity the turbulence is supposed to be homogeneous and isotropic:

$$\langle V \rangle^0 = V_0 = \mathrm{const}, \quad \langle \mathbf{u} \rangle^0 = 0, \quad \langle p \rangle^0 = \mathrm{const}, \quad \langle u'_i u'_k \rangle^0 = c^2 d_{ik}, \qquad (2.2)$$

where $c = \mathrm{const}$. We consider the case of the low-pressure high-energy turbulence

$$\langle p \rangle^0 \ll V_0 c^2, \qquad (2.3)$$

and

$$\langle p \rangle \geq 0.$$

This system will be taken as a substratum for modeling the physical fields and particles [2].

## 3 Reynolds equations

The motion of an inviscid fluid is considered. The fluid density $V(\mathbf{x},t)$ is supposed to do not fluctuate

$$V=\langle V\rangle. \tag{3.1}$$

Putting (2.1), (3.1) in the Euler equation

$$\partial_t u_i + u_k \partial_k u_i + \frac{1}{V}\partial_i p = 0, \tag{3.2}$$

averaging and taking account of $\langle \mathbf{u}'\rangle=0, \langle p'\rangle=0$, we find for turbulence averages:

$$\partial_t\langle u_i\rangle + \langle u_k\rangle\partial_k\langle u_i\rangle + \langle u'_k \partial_k u'_i\rangle + \frac{1}{\langle V\rangle}\partial_i\langle p\rangle = 0. \tag{3.3}$$

Here and further on $\partial_t = \partial/\partial t$, $\partial_k = \partial/\partial x_k$, $i,k=1,2,3$ and summation over recurrent index is implied throughout. Next, we suppose that the fluctuation component $\mathbf{u}'$ relates to a solenoidal velocity field of the vortex turbulence i.e.

$$\partial_k u'_k = 0.$$

Using the latter in (3.3) gives

$$\partial_t\langle u_i\rangle + \langle u_k\rangle\partial_k\langle u_i\rangle + \partial_k\langle u'_i u'_k\rangle + \frac{1}{\langle V\rangle}\partial_i\langle p\rangle = 0. \tag{3.4}$$

Equation (3.4) represents the first link in the chain of dynamical equations for consecutive moments of turbulent pulsations. Next equation is received multiplying (3.2) by $u'_l$, symmetrizing, substituting in it (2.1), (3.1) and averaging:

$$\langle u'_k\left[\partial_t(\langle u_i\rangle+u'_i)+(\langle u_j\rangle+u'_j)\partial_j(\langle u_i\rangle+u'_i)+\frac{1}{\langle V\rangle}\partial_i(\langle p\rangle+p')\right]+$$

$$u'_i\left[\partial_t(\langle u_k\rangle+u'_k)+(\langle u_j\rangle+u'_j)\partial_j(\langle u_k\rangle+u'_k)+\frac{1}{\langle V\rangle}\partial_k(\langle p\rangle+p')\right]\rangle = 0. \tag{3.5}$$

An infinite chain of Reynolds equations provides us with a complete description of the averaged turbulence.

Integrating (3.4) for the case of the isotropic turbulence in an incompressible medium, one may get:

$$\langle u'_i u'_k\rangle = \langle u'_1 u'_1\rangle d_{ik},$$

$$V_0\langle u'_1 u'_1\rangle + \langle p\rangle = V_0 c^2 + \langle p\rangle^0. \tag{3.6}$$

This is a kind of Bernoulli equation and actually an equation of state of the ideal isotropic turbulence. It implies rather a broad range of variation for the turbulence energy density $1/2\langle u'_i u'_i \rangle$ and pressure, involving coexistence of different turbulence phases.

## 4 Perturbations of the turbulence

We are interested in deviations of the turbulence from the background (2.2):
$$\boldsymbol{d}\langle \mathbf{u}\rangle = \langle \mathbf{u}\rangle - \langle \mathbf{u}\rangle^0, \quad \boldsymbol{d}\langle \boldsymbol{p}\rangle = \langle \boldsymbol{p}\rangle - \langle \boldsymbol{p}\rangle^0, \quad \boldsymbol{d}\langle u'_i u'_k\rangle = \langle u'_i u'_k\rangle - \langle u'_i u'_k\rangle^0$$
and
$$\boldsymbol{d}\langle \boldsymbol{V}\rangle = \langle \boldsymbol{V}\rangle - \langle \boldsymbol{V}\rangle^0.$$
When
$$\left|\boldsymbol{d}\langle \boldsymbol{p}\rangle\right| \leq \langle \boldsymbol{p}\rangle^0,$$
we have in the low-pressure substratum (2.3):
$$\left|\boldsymbol{d}\langle \boldsymbol{p}\rangle\right| \ll V_0 c^2, \quad \left|\boldsymbol{d}\langle u'_i u'_k\rangle\right| \ll c^2, \quad \left|\boldsymbol{d}\langle u\rangle\right| \ll c, \qquad (4.1)$$

where (3.6) has been used for the evaluation. As will be shown below, this also entails
$$\left|\boldsymbol{d}\langle \boldsymbol{V}\rangle\right| \ll V_0. \qquad (4.2)$$
With the small perturbations of the turbulence, the Reynolds equations can be linearized [1]. Denoting
$$p = \boldsymbol{p}/V_0, \qquad (4.3)$$
we have for (3.4) and (3.5), respectively
$$\partial_t \boldsymbol{d}\langle u_i\rangle + \partial_k \boldsymbol{d}\langle u'_i u'_k\rangle + \partial_i \boldsymbol{d}\langle p\rangle = 0, \qquad (4.4)$$
$$\partial_t \boldsymbol{d}\langle u'_i u'_k\rangle + c^2\left(\partial_i \boldsymbol{d}\langle u_k\rangle + \partial_k \boldsymbol{d}\langle u_i\rangle\right) + h_{ik} = 0, \qquad (4.5)$$
where
$$h_{ik} = \left(\langle u'_i \partial_k p'\rangle + \langle u'_k \partial_i p'\rangle\right) + \partial_j \langle u'_i u'_j u'_k\rangle.$$
Here $c$ takes the sense of the speed for the wave of turbulence perturbation.

## 5  Maxwell's equations

Differentiating (4.5) with respect to $x_k$, we get the vector equation

$$\partial_t \partial_k \boldsymbol{d}\langle \mathbf{u}' u'_k \rangle - c^2 \nabla \times \nabla \times \boldsymbol{d}\langle \mathbf{u}\rangle + \mathbf{g} = 0, \tag{5.1}$$

where

$$g_i = \partial_k h_{ik} + 2c^2 \partial_i \nabla \cdot \boldsymbol{d}\langle \mathbf{u}\rangle$$

and the identity $\nabla(\nabla \cdot) = \nabla \times \nabla \times + \nabla^2$ was taken advantage of. The terms of (5.1) are of the same order of magnitude as those in (4.4).

With the definitions [1]

$$A_i = \boldsymbol{k} c \boldsymbol{d}\langle u_i \rangle, \tag{5.2}$$

$$\boldsymbol{j} = \boldsymbol{k} \boldsymbol{d}\langle p \rangle, \tag{5.3}$$

$$E_i = \boldsymbol{k} \P_k \boldsymbol{d}\langle u'_i u'_k \rangle,$$

$$j_i = \frac{\boldsymbol{k}}{4\boldsymbol{p}} g_i,$$

where $\boldsymbol{k}$ is an arbitrary constant, (4.4) and (5.1) take the appearance of Maxwell's equations, respectively,

$$\frac{1}{c}\frac{\partial \mathbf{A}}{\partial t} + \mathbf{E} + \nabla \boldsymbol{j} = 0,$$

$$\frac{1}{c}\frac{\partial \mathbf{E}}{\partial t} - \nabla \times (\nabla \times \mathbf{A}) + \frac{4\boldsymbol{p}}{c} \mathbf{j} = 0,$$

It is instructive to express the plane electromagnetic wave in terms of $\boldsymbol{d}\langle u'_i u'_k \rangle$. We get for a continuous incompressible substratum:

$$h_{ik} = 0, \quad \boldsymbol{d}\langle p \rangle = 0,$$

$$\boldsymbol{d}\langle u_2 \rangle = l_2 F(\boldsymbol{w} t - k_1 x_1),$$

$$\boldsymbol{d}\langle u'_1 u'_2 \rangle = \frac{2c^2 k_1 l_2}{\boldsymbol{w}} F(\boldsymbol{w} t - k_1 x_1).$$

So, in this process the density of the turbulence energy $1/2\langle u'_i u'_i \rangle$ remains unperturbed:

$$\boldsymbol{d}\langle u'_1 u'_1 \rangle = 0.$$

# 6  The Lorentz gauge

The so-called gauge transformation is concerned with a kinematical part of the theory. It is originated from the mass balance in the medium. Putting (2.1), (3.1) in the continuity equation

$$\partial_t V + \nabla \cdot (V\mathbf{u}) = 0$$

and averaging, adds to the Reynolds equations:

$$\partial_t \langle V \rangle + \nabla \cdot (\langle V \rangle \langle \mathbf{u} \rangle) = 0. \tag{6.1}$$

We have for small variations of the pressure

$$d\langle p \rangle = b^2 d\langle V \rangle,$$

where $b$ is the speed of a density wave in the medium. The supposition is that the wave of turbulence perturbation propagates as a compound density-turbulence perturbation wave i.e.

$$b = c.$$

Therefore, it can be taken

$$d\langle p \rangle = c^2 d\langle V \rangle. \tag{6.2}$$

(From the thermodynamical point of view, in this system $c^2 = kT$. So, formally (6.2) takes the sense of the ideal gas equation.) With (6.2) the inequality (4.2) appears to be the consequence of the first inequality in (4.1)

Then, linearizing the averaged continuity equation (6.1), we get

$$\partial_t d\langle V \rangle + V_0 \nabla \cdot d\langle \mathbf{u} \rangle = 0.$$

Substituting to this (6.2) gives in terms of the specific pressure (4.3)

$$\partial_t d\langle p \rangle + c^2 \nabla \cdot d\langle \mathbf{u} \rangle = 0.$$

With the definitions (5.2), (5.3) the latter takes the appearance of the Lorentz gauge:

$$\frac{1}{c}\frac{\partial j}{\partial t} + \nabla \cdot \mathbf{A} = 0.$$

So, validity of the Lorentz gauge may indicate that the electrostatic field is accompanied by the slight variation of the substratum density. On this account, the scattering of a neutral particle by the electrostatic field should be expected.

The density-perturbation wave of the turbulence medium can serve as a model of the photon.

## 7 Caviton

We are interested in stationary perturbations of the turbulence. Let us consider the case of an incompressible medium: $\partial_i \langle u_i \rangle = 0$. Assuming that in the continuous medium $h_{ik} = 0$ and taking in (4.5) $k = i$, we get:

$$\partial_t d \langle u'_i u'_i \rangle = 0.$$

So, for linear perturbations of the turbulence in a continuous incompressible medium the energy profile is conserved. However, in general the medium is discontinuous.

Let we have an empty bubble in an incompressible fluid. The medium is taken to be noncorpuscular. So, it will not fill the bubble with a vapor in order to attain the equilibrium. Instead, the turbulence is perturbed reaching at the wall of the bubble

$$\langle u'_1 u'_1 \rangle_R = c^2 + \langle p \rangle^0,$$

where $\langle p \rangle^0 = \langle \mathbf{p} \rangle^0 / V_0$. This perturbation is found from (3.6), taking in it the equilibrium pressure

$$\langle \mathbf{p} \rangle_R = 0.$$

Thus, the cavity is stabilized forming a *caviton*. A caviton occurs as a centre, or source of the stationary perturbation field. As was shown [2], the latter has the form of the Coulomb field. With account of the boundary condition, we have outside the core of the stable cavity:

$$d \langle u'_1 u'_1 \rangle = \frac{\langle p \rangle^0 R}{|\mathbf{x} - \mathbf{x}'|}, \qquad (7.1)$$

where $\mathbf{x}'$ is the center and $R$ the radius of the cavity (Fig.1, top). This is the model of the proton and the electrostatic field generated by it.

The energy of the turbulence perturbation is given by

$$dU = 1/2 \int V d \langle u'_i u'_i \rangle d^3 x, \qquad (7.2)$$

where $d^3 x = dx_1 dx_2 dx_3$. It is infinite for the Coulomb field.

The neutron is modeled by a non-equilibrium cavity in the fluid (Fig.3). In experiments we have

$$n \rightarrow p^+ + e^- + \tilde{n}.$$

So, the positive perturbation energy due to the proton must be counterbalanced by the negative perturbation energy due to the electron. On this account, we take for the electron an islet of the quiescent fluid. It generates the perturbation field

$$d\langle u_1' u_1' \rangle = -\frac{c^2 r_e}{|\mathbf{x} - \mathbf{x}'|}, \qquad (7.3)$$

where $r_e$ is the radius of the core (Fig.6, top). The condition of mutual compensation of the two infinite quantities – (7.2) with (7.1) and (7.2) with (7.3) – gives

$$c^2 r_e = \langle p \rangle^0 R.$$

So, in the high-energy low-pressure substratum (2.3)

$$r_e \ll R.$$

The sum of the two infinite quantities under consideration is not vanishing, but gives a finite quantity. It relates to the energy of the neutrino.

## 8 Particles and antiparticles

In this section, the caviton models of the particles are extended to the case of a compressible substratum. For our purposes, it is sufficient to consider here only the hollow-bubble models of the particles and respective models of the antiparticles, or vice versa.

The model of the proton is shown in Fig.1. The model of the antiproton (Fig.2) is easily obtained by mirroring the graphs of Fig.1 about the respective asymptotes. So, the antiproton is modeled by the inclusion of a drop of the lowered-energy fluid. The qualitative correctness of these models can be verified using it in order to interpret the reaction of annihilation:

$$\text{particle} + \text{antiparticle} \rightarrow \text{photons} + (\text{neutrinos} + \text{antineutrinos}). \qquad (8.1)$$

The energy density of the perturbation, generated by a particle, is exactly opposite to that of the respective antiparticle. We have for the proton

$$dU_+ = +M + \frac{4\pi}{3} R^3 V_0 \frac{3}{2} \langle p \rangle^0$$

and for the antiproton

$$dU_- = -M - \frac{4\pi}{3} R^3 V_0 \frac{3}{2} \langle p \rangle^0.$$

After the annihilation

$$dU = dU_+ + dU_- = 0.$$

The electromagnetic energy is given by

$$e = \frac{k^2}{8\pi} \int \left( \partial_k d \langle \mathbf{u}' u_k' \rangle \right)^2 d^3 x.$$

We have
$$e_+ = e_- > 0.$$
After annihilation
$$e = e_+ + e_- = 2e_+.$$
This finite quantity corresponds to photons.

Until this moment, the model was identical to that in the incompressible substratum [2]. Next, we discuss the features, which are added to it by the compressibility. For the substratum density we have
$$d\langle V\rangle_+ = -d\langle V\rangle_-$$
(see Fig.1, bottom and Fig.2, bottom). After annihilation:
$$\int d\langle V\rangle d^3 x = 0.$$
If we take for the particle's mass
$$m = \int f\left(\langle V\rangle - \langle V\rangle^0\right) d^3 x,$$
where $f(-d\langle V\rangle) = -f(d\langle V\rangle)$, then
$$m_+ + m_- = 0.$$

The neutrinos + antineutrinos, which can be formed in the reaction (8.1), are supposed to compensate each other by the same scheme.

The hollow-bubble model of the neutron and the model of the antineutron are shown in Fig.3 and Fig.4, respectively.

The model of the localized electron is shown in Fig.6, the respective model of the positron is shown in Fig.5. You see that here perturbations of turbulence are far beyond (4.1). So, a delocalization until $d\langle p\rangle \leq \langle p\rangle^0$ was considered [2]. Following this line, in Fig.8 a $1/N$-th splinter of the electron is shown, where $N = c^2/\langle p\rangle^0$. The respective splinter of the positron is shown in Fig.7. Notice that the radius $r_e$ of the splinter's core equals to that of the whole electron, whereas the perturbation fields are $N$ times lower.

Qualitatively, the positron (Fig.7) has the same features as that of the proton (Fig.1), and the electron (Fig.8) – the same as the antiproton (Fig.2). However, the hollow-bubble model gives a sharp jump of the density at the boundary of the nucleon's core. This may serve as an indication of the complex inner structure of the nucleon. No jump of density in the electron indicates "the absence" of the internal structure. Analogously, Fig.3 and Fig.4 express the features of the neutrino and antineutrino as well.

We do not know what is the maximal value of the density. If it is about $2\langle V\rangle^0$, then the isle of the quiescent fluid, which represents a localized electron, may just be the medium free of an admixture of the empty space.

## 9  Order through chaos – a microscopic view

We can not derive the self-organization in the turbulent medium. Nor can we prove its stability. The system is too complicated to handle it *ab initio*. Still, it can be approached with some additional supposition. First, the turbulence was supposed to have a vortex microscopic structure. Then, a nonlinear Schroedinger equation can be applied [4]. The model was shown numerically [5] to be attracted to a soliton solution.

A heuristic microscopic model of the turbulent ideal fluid with voids has been first proposed by John Bernoulli Jr. more than two centuries ago. It is known historically as the vortex sponge. We have two basic vortex configurations: the straight vortex tube and the closed vortex tube (vortex ring). Hence, the two basic kinds of the vortex sponge can be imagined: the random heap of the vortex tubes [3] and the packing of the vortex rings [6]. The vortices are supposed to be hollow inside. So, to redistribute the empty space in the fluid may mean just to rearrange the tubes. Here, we take for the vortex sponge the heap of randomly oriented straight vortex tubes [3]. Then, a particle may correspond to a closed vortex formation involving in itself the empty space. For instance, when a fluid ball moves in the fluid it generates the velocity field of a vortex ring. When a bubble moves in the fluid it must generate the velocity field which is opposite to the latter. Such can be the vortex structure of the neutrino and antineutrino and also that of the moving neutron or antineutron. A loop on the vortex tube may represent the neutron at rest [4]. A torsional (helical or kink) wave on a vortex filament models the electromagnetic wave. We may also have axisymmetric (area-varying) waves propagating along a vortex tube [7]. This can be taken for the model of the gravitational wave. Macroscopically, it should be described as the flow of the transient point dilatation.

Clearly, discreteness of the particles and charges are determined by the discreteness of the vortex sponge. In the average, it is defined by $L$ – the mean length of the vortex filaments per a unit of volume. The vortex sponge possesses the elastic properties that we just appreciate as electromagnetism. In this context, the above reported model should be look at only as a mesoscopic description of the microscopic structure.

This description can be adjusted to some new microscopic features without changing the general picture. For instance, if the vortex structures modeling the neutron and proton include in their cores some portions of the fluid, this can be accounted for in the mesoscopic model taking for minimal density and pressure the values which are different from zero: $\langle V \rangle \geq V_m$, $\langle p \rangle \geq p_m$. So, concerning their cores, the particle models presented in Figures 1-4 there quite can be found to be some exaggeration of the real situation.

## 10  Conclusion

We have shown that in mechanical models of electromagnetism the Coulomb gauge corresponds to an incompressible substratum. Validity of the Lorentz gauge points out to that the substratum can be effectively compressible. The latter should be due to volume dispersion of the empty space in the microscopically incompressible fluid. The possibility for the substratum density to have both negative and positive deviations from a background level makes conditions for generating particle-antiparticle pairs from the background.

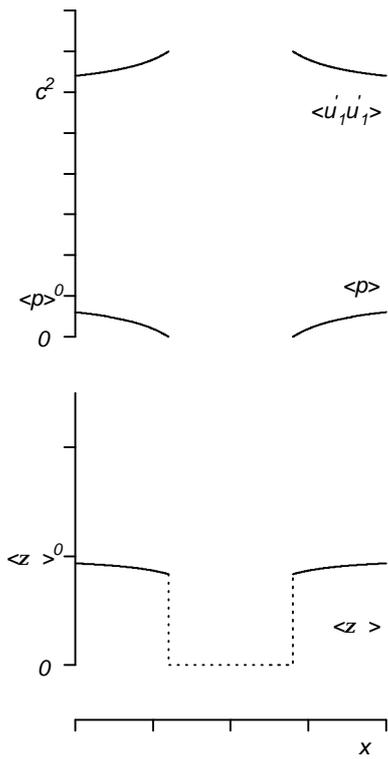 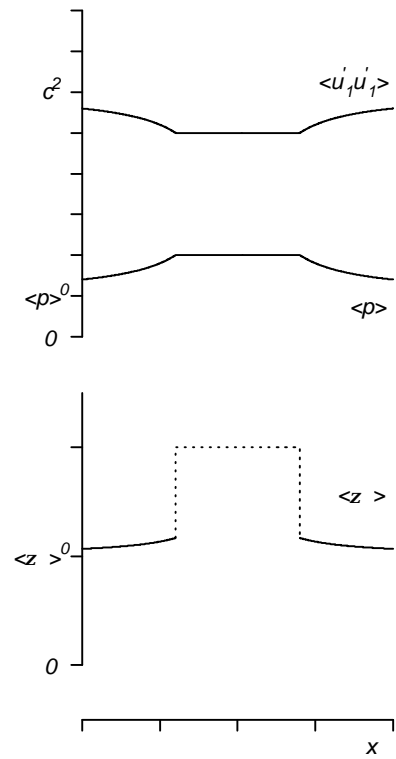

Fig. 1                                                 Fig. 2

The hollow-bubble model of the proton.         The antiproton.

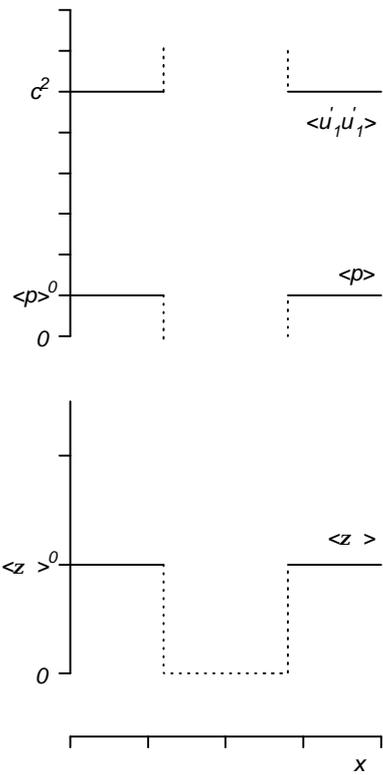 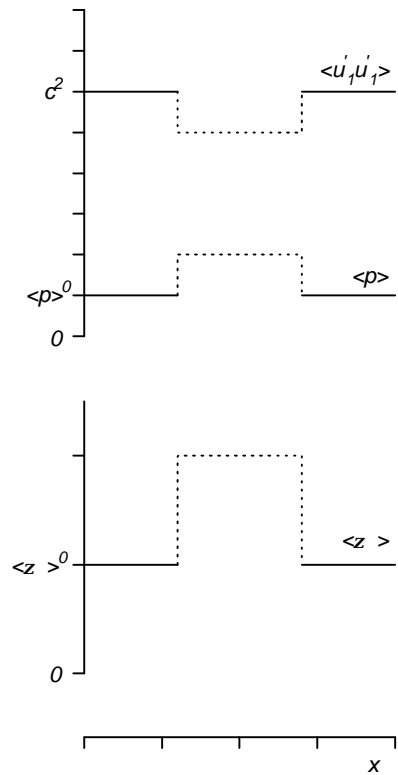

Fig. 3                                     Fig.4

The hollow-bubble model of the neutron.       The antineutron.

.

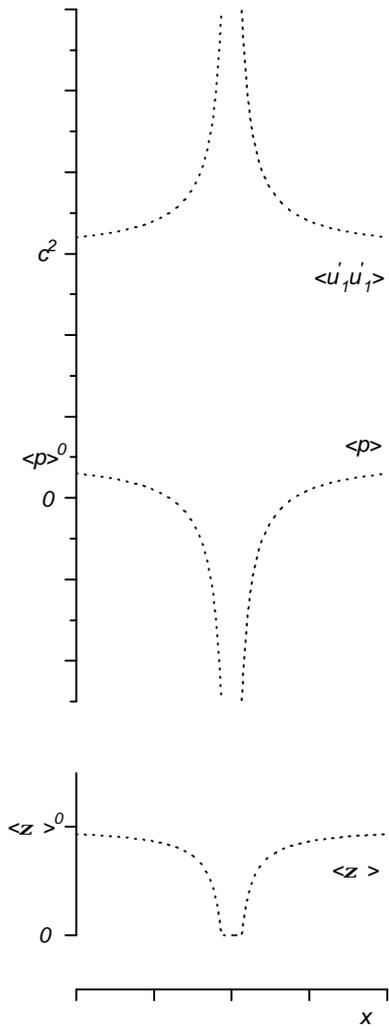 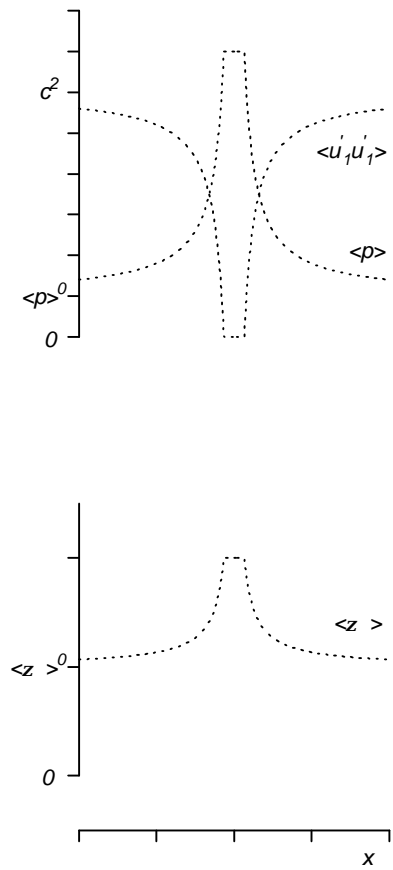

Fig.5 Fig.6

The localized positron. The localized electron.

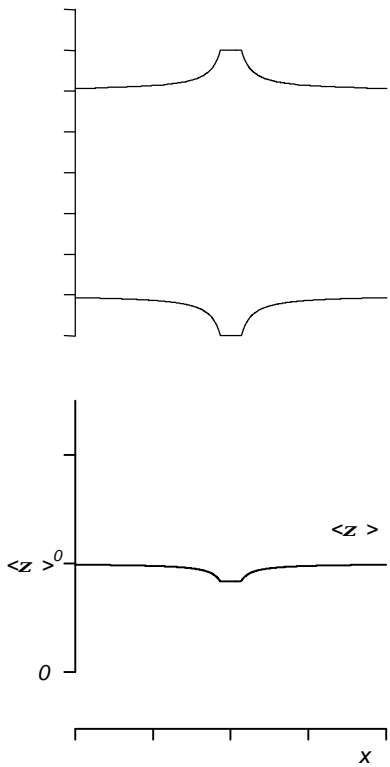 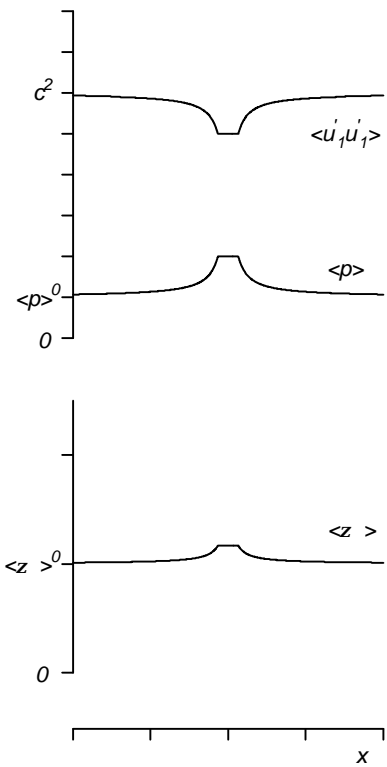

Fig. 7.    Fig. 8.

The $1/N$-th splinter of the positron.    The $1/N$-th splinter of the electron.

$$N = c^2 / \langle p \rangle^0. \text{ Here } N=6.$$

Bologna, March 26, 1999

Dr. V.P.DMITRIYEV
P.O.Box 160
Moscow 117574
Russia

Our Ref.  8106/NCAT

Dear Doctor Dmitriyev

we are sorry to inform you that your paper *"Mechanical analogies of the Lorentz gauge, particles and antiparticles"* has been rejected.

Please find enclosed the referee's report.

Yours sincerely

The Vice-Director
Renato A.Ricci

**Referee's Report on the paper 8106/NCAT  "Mechanical analogies for the Lorentz gauge, particles and antiparticles"** – by V.P.Dmitriyev

The paper recently published by the author (Nuov.Cim. 111A, 501–511, 1998) presents a mechanical analogy of particles and fields with a turbulent fluid. I consider this description of some interest, but certainly it cannot be taken as a realistic theory of particles and interactions; therefore, I do not find a real motivation for pushing the analogy beyond the general picture already given.
On my opinion the present contribution does not add new information of physical interest to justify the publication on Nuovo Cimento A.

16